\documentclass[]{interact}

\usepackage{epstopdf}% To incorporate .eps illustrations using PDFLaTeX, etc.
\usepackage[caption=false]{subfig}% Support for small, `sub' figures and tables

\usepackage{natbib}% Citation support using natbib.sty
\bibpunct[, ]{(}{)}{;}{a}{}{,}% Citation support using natbib.sty
\renewcommand\bibfont{\fontsize{10}{12}\selectfont}% Bibliography support using natbib.sty

\usepackage{multirow}
\usepackage{diagbox}
\usepackage[noabbrev,capitalize]{cleveref}
\usepackage{xcolor}

\newcommand{\myparagraph}[1]{%
\vspace{0.5em}
\noindent\textbf{#1.}
}

\theoremstyle{plain}% Theorem-like structures provided by amsthm.sty

\theoremstyle{definition}

\theoremstyle{remark}

\ifdefined\showrevision
\usepackage{xcolor}
\usepackage[normalem]{ulem}
\usepackage{marginnote}

\newcommand{\revref}[2]{%
\marginnote{$R_{#1}C_{#2}$}
}
\newcommand{\revmod}[1]{%
{\color{blue}#1}
}

\else
\newcommand{\marginnote}[1]{\ignorespaces}
\newcommand{\revref}[2]{\ignorespaces}
\newcommand{\revmod}[1]{#1}

\newcommand{\revdel}[1]{}
\fi
\begin{document}

%\articletype{ARTICLE TEMPLATE}% Specify the article type or omit as appropriate

%\title{Hyperspectral Image Demosaicking and Spectral Correction for Medical Images with Supervised Learning and Synthetic Snapshot Mosaic Datasets}
\title{Deep Learning Approach for Hyperspectral Image Demosaicking, Spectral Correction and High-resolution RGB Reconstruction}

\author{
\name{Peichao Li\textsuperscript{a}\thanks{Contact: Peichao Li -- Email: peichao.2.li@kcl.ac.uk},
%and 
Michael Ebner\textsuperscript{a,b},
%and 
Philip Noonan\textsuperscript{a},
%and 
Conor Horgan\textsuperscript{a,b},
%and 
Anisha Bahl\textsuperscript{a},
%and 
S\'ebastien Ourselin\textsuperscript{a,b},
%and 
Jonathan Shapey\textsuperscript{a,b,c}
and 
Tom Vercauteren\textsuperscript{a,b}}
\affil{
\textsuperscript{a}School of Biomedical Engineering \& Imaging Sciences, King's College London, London, UK\\
\textsuperscript{b}Hypervision Surgical Ltd, London, UK\\
\textsuperscript{c}Department of Neurosurgery, King's College Hospital NHS Foundation Trust, London, UK
}
}

\maketitle

\begin{abstract}
Hyperspectral imaging is one of the most promising techniques for intraoperative tissue characterisation. Snapshot mosaic cameras, which can capture hyperspectral data in a single exposure, have the potential to make a real-time hyperspectral imaging system for surgical decision-making possible. However, optimal exploitation of the captured data requires solving an ill-posed demosaicking problem and applying additional spectral corrections to recover spatial and spectral information of the image. In this work, we propose a deep learning-based image demosaicking algorithm for snapshot hyperspectral images using supervised learning methods. Due to the lack of publicly available medical images acquired with snapshot mosaic cameras, a synthetic image generation approach is proposed to simulate snapshot images from existing medical image datasets captured by high-resolution, but slow, hyperspectral imaging devices. Image reconstruction is achieved using convolutional neural networks for hyperspectral image super-resolution, followed by cross-talk and leakage correction using a sensor-specific calibration matrix. The resulting demosaicked images are evaluated both quantitatively and qualitatively, showing clear improvements in image quality compared to a baseline demosaicking method using linear interpolation. Moreover, the fast processing time of~45\,ms of our algorithm to obtain super-resolved RGB or oxygenation saturation maps per image frame for a state-of-the-art snapshot mosaic camera demonstrates the potential for its seamless integration into real-time surgical hyperspectral imaging applications.
\end{abstract}

%\begin{keywords}\todo{}
%hyperspectral imaging; demosaicking; deep learning; surgical imaging
%\end{keywords}

\section{Introduction}
Reliable discrimination between tumour and surrounding tissues remains a challenging task in surgery and in particular in neuro-oncology surgery.
Despite intensive research and progress in advanced computer-assisted visualisation techniques, most intraoperative surgical evaluations are still heavily reliant on subjective visual assessment from clinicians. 
Modern intraoperative tissue discrimination techniques therefore often involve the use of interventional techniques such as fluorescence and ultrasound imaging.
%Such approaches typically disrupt the clinical workflow and interpretation of intraoperative information remains highly subjective therefore hindering accurate surgical operations.
\revref{1}{1}
\revmod{However, visual assessment of fluorescence intensities during surgeries are usually qualitative, which hinders accurate, reliable and repeatable measurements for consensus, standardization, and adoption of fluorescence-guided surgery in the field~\citep{valdes2019fluorescence}. 
Ultrasound imaging may suffer from poor resolution and restricted field of view, and the interpretation is highly subjective to experience from the experts~\citep{aingaya2021ultrasound}.}

\begin{figure}[tb!]
\centering
\subfloat[Hyperspectral cube]{%
\resizebox*{6cm}{!}{\includegraphics{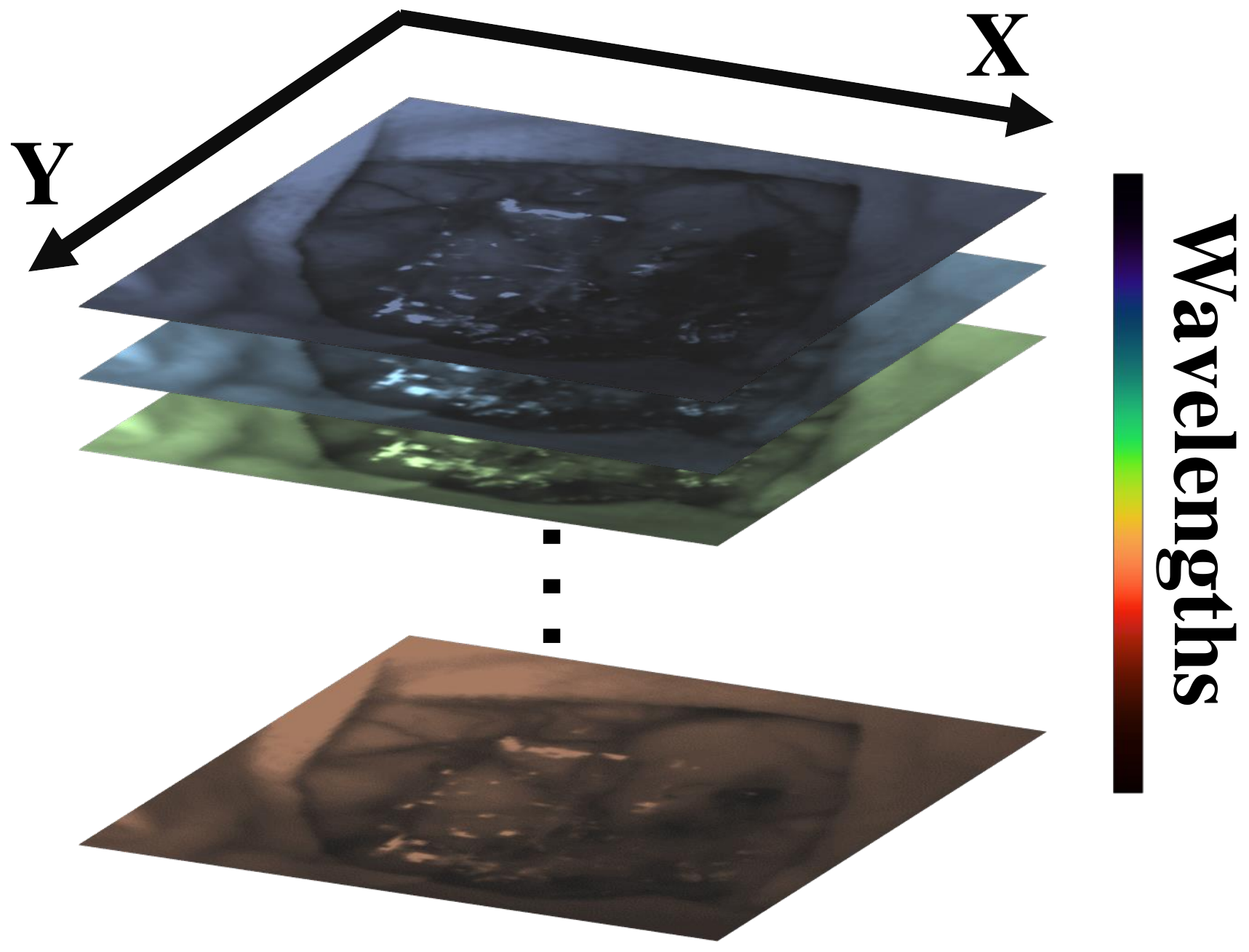}}}\hspace{25pt}
\subfloat[Subsampling and demosaicking operators]{%
\resizebox*{6cm}{!}{\includegraphics{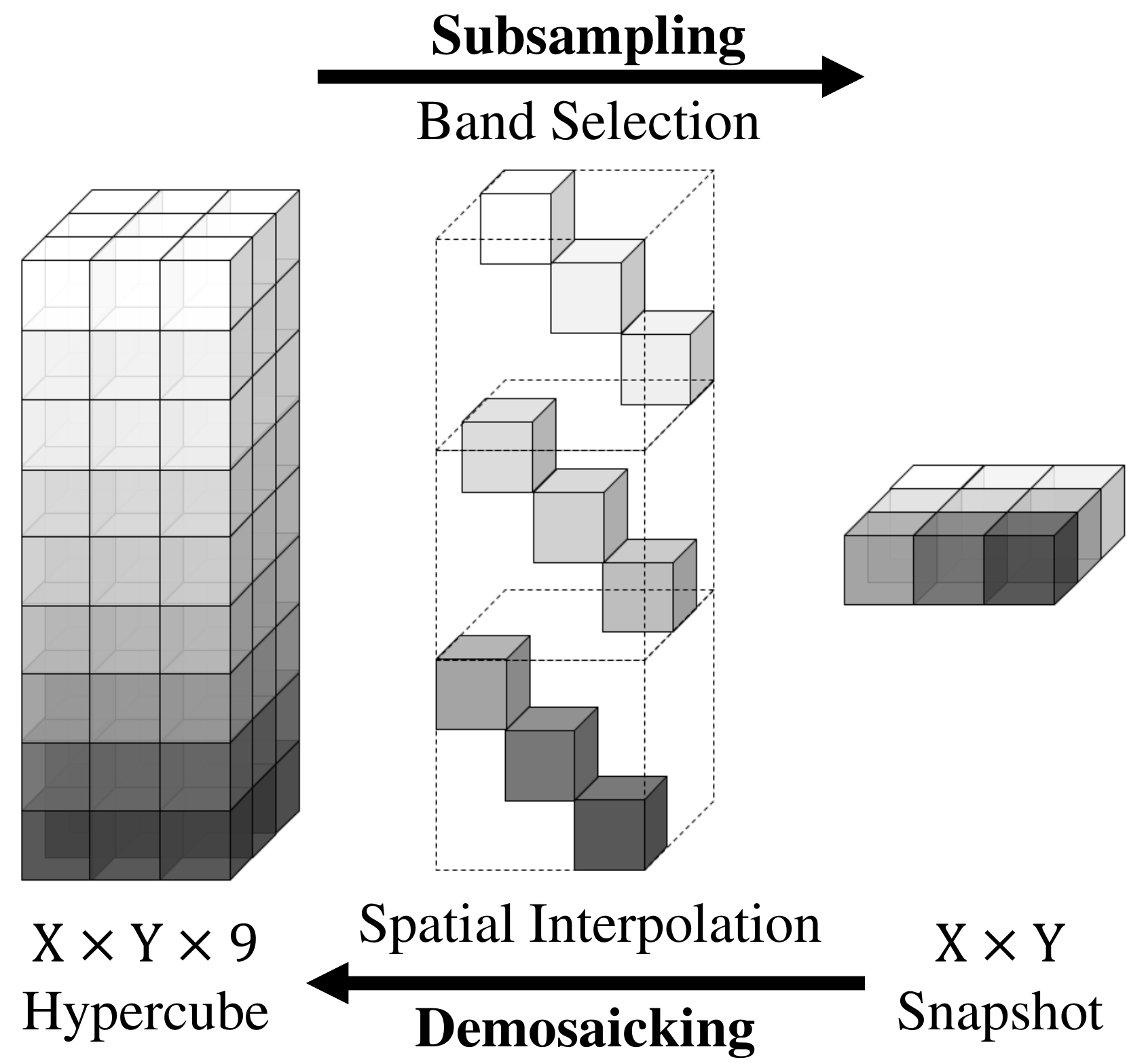}}}
\caption{Examples to illustrate a hyperspectral cube as well as subsampling and demosaicking operations: (a)~shows the spatial dimensions (X and Y) and spectral dimension of a hypercube;
(b)~shows how hyperspectral cube and snapshot mosaic images can be transformed into each other with band selection/spatial interpolation. Due to the space constraint of the image, $3 \times 3$ snapshot mosaicking is taken as an example.} \label{hypercube-mosaic}
\end{figure}

Intraoperative hyperspectral imaging (HSI) provides a non-contact, non-ionising and non-invasive solution suitable for many medical applications~\citep{lu2014medical,shapey2019intraoperative,clancy2020surgical}.
HSI can provide rich high-dimensional spatio-spectral information within the visible and near-infrared electromagnetic spectrum across a wide field of view.
Compared to conventional colour imaging that provides red, green, and blue (RGB) colour information, HSI can capture information across multiple spectral bands beyond what the human eye can see therefore facilitating tissue differentiation and characterisation.
\revref{1}{1}
\revmod{Unlike fluorescence and ultrasound imaging, HSI exploits the inherent optical characteristics of different tissue types. It captures the measurements of light which provide quantitative diagnostic information on tissue perfusion and oxygen saturation, enabling improved tissue characterisation relative to fluorescence and ultrasound imaging~\citep{lu2014medical}.}
Depending on the number of acquired spectral bands, hyperspectral imaging may also be referred to as multispectral imaging, but for simplicity the hyperspectral terminology will be used.
A single hyperspectral image data typically spans three dimensions, two of them represent 2D spatial dimensions and the other represents spectral wavelengths, as illustrated in \cref{hypercube-mosaic}(a). 
Therefore, 3D HSI data are thus often referred to as hyperspectral cubes, or hypercubes in short.
In addition, time comes as a fourth dimension in the context of dynamic scenes such as those acquired during surgery.
Hyperspectral cameras can broadly be divided into three categories based on their acquisition methods, namely spatial scanning, spectral scanning and snapshot cameras \citep{shapey2019intraoperative,clancy2020surgical}.
Spatial scanning acquires the entire wavelength spectrum simultaneously on either a single pixel or a line of pixels using linear or 2D array detector, respectively.
The camera will spatially scan through pixels over time to complete the hyperspectral cube capturing. Spectral scanning, on the other hand, is able to capture the entire spatial scene at a certain wavelength with a 2D array detector, and then switches to different wavelengths over time to complete scanning.
These two types of spectral cameras are able to acquire hyperspectral data with high spatial and spectral resolution, but long acquisition times prevent them from providing live image displays suitable for real-time intraoperative use. 

To achieve intraoperative tissue characterization with HSI in real-time, snapshot cameras are more suitable as they can capture hyperspectral cube data in real-time \citep{ebner2021intraoperative}.
A common type of snapshot camera uses a snapshot mosaic system to acquire the entire hyperspectral cube instantly without the need of a scanning mechanism.
The refined $n \times n$ pixel filter array, arranged similarly to the $2 \times 2$ colour filter array on the RGB sensor, allows the snapshot camera to acquire a maximum of $n^2$ different spectral bands in a single exposure~\citep{geelen2014compact}.
Other snapshot hyperspectral imaging approaches such as coded aperture snapshot spectral imaging (CASSI)~\citep{wagadarikar2008single} and micro-lens based acquisition have been proposed. 
In general, the downside of snapshot acquisition is that it sacrifices spatial and spectral resolution to achieve fast data acquisition speeds. \cref{hypercube-mosaic}(b) illustrates the relationship between a high-resolution hyperspectral cube and a $3 \times 3$ snapshot mosaic image as a simplified example. An $X \times Y$ snapshot image is composed of a large number of individual $3 \times 3$ tiles following mosaic patterns. The $3 \times 3$ snapshot on the right of \cref{hypercube-mosaic}(b) is an example of a single tile captured by the $3 \times 3$ sensor array. 

As the image captured by a snapshot mosaic sensor is in~2D, a demosaicking operation is necessary to restore the spatial and spectral resolution of the image, followed by spectral correction to deal with the parasitic effects of the sensors, such as harmonics, cross-talks and leakage \citep{pichette2017calibration}. 
Spectral correction can be usually handled by applying a calibration matrix such as provided by the camera manufacturer, but demosaicking of the snapshot data is challenging. 
As illustrated in \cref{hypercube-mosaic}(b), the demosaicking operator usually involves splitting the image into different spectral bands, followed by spatial interpolation to fill in the missing data. Common ways of image demosaicking using interpolation methods usually result in poor image quality of reconstructed hyperspectral data, so several approaches have been presented to address this demosaicking problem. For example, Hy-Demosaicing proposed by Zhuang et al. used data-adaptive subsampled signal subspaces for reconstruction of hyperspectral urban images by exploiting the low-rank and self-similarity properties of the hyperspectral images \citep{zhuang2018hydemosaicing}. Deep learning methods for hyperspectral demosaicking were also investigated, such as the similarity maximisation framework proposed for performing end-to-end demosaicking and cross-talk correction for agricultural machine vision \citep{dijkstra2019hyperspectral}. 

Despite the development of demosaicking algorithms, research on medical hyperspectral image demosaicking remains limited. The goal of this study is to develop a reliable real-time image demosaicking, spectral correction and associated RGB reconstruction algorithm to recover higher quality medical hyperspectral images suitable for intraoperative applications.
Due to the lack of open datasets of snapshot mosaic hyperspectral imaging from intraoperative settings, and more importantly, due to the impossibility of capturing hyperspectral imagery paired for both snapshot and high-resolution sensors,
the proposed learning-based demosaicking algorithm makes use of publicly available medical hyperspectral image datasets captured in high spatial and spectral resolution by line-scan cameras for training purposes.
From high-resolution data, we exploit the knowledge of the physical image acquisition process to simulate images expected from a snapshot mosaic camera as well as their corresponding ideal demosaicked images.
This allows us to form image pairs suitable for supervised training. The results have been evaluated with popular full-reference image quality metrics including structural similarity (SSIM) and peak signal-to-noise ratio (PSNR). A first qualitative survey has also been conducted on the reconstructed RGB image quality, and the proposed algorithm has been applied to real snapshot mosaic test images to demonstrate its effectiveness. The speed and quality of the reconstructed image from our proposed algorithm showed respectable results, which will facilitate seamless integration into intraoperative hyperspectral imaging systems using snapshot mosaic cameras for responsive surgical guidance~\citep{ebner2021intraoperative}.

\section{Material and Methods}
%
\iffalse % commenting this out for the sake of space
One of the major challenges for developing learning-based hyperspectral image demosaicking algorithms is the lack of hyperspectral datasets offering paired snapshot and high-resolution data. Such datasets would be even more complex to acquire in intraoperative contexts.
We took an alternative approach where synthetic low-resolution snapshot images are generated from higher-resolution hyperspectral images captured by line-scan sensors endowed with long acquisition times.
Our developed demosaicking algorithm can thus take advantage on the resulting synthetic paired high-resolution/snapshot data.
\fi
This section first introduces the publicly available hyperspectral line-scan datasets used in the experiments. Next, the overall framework for simulating the snapshot image acquisition process using the line-scan data will be presented, with details on how synthetic snapshot images and ideal demosaicked images are generated. After that, this section introduces the integration of supervised image super-resolution methods into the demosaicking, spectral correction and RGB generation framework.

\subsection{Source Datasets}
Line-scan sensors are able to capture data across hundreds of spectral bands within the visible and near-infrared range.
While they require long acquisition times, they provide high spatial and spectral resolution.
Line-scan data contains sufficient information for generating snapshot mosaic images which have much lower spatial and spectral resolution. Two publicly available line-scan hyperspectral image datasets have been used in this work and are presented hereafter.

\cite{fabelo2019helicoid} provide a hyperspectral dataset acquired during neurosurgical procedures as part of the HypErspectraL Imaging Cancer Detection (HELICoiD) project.
This dataset contains 36 hyperspectral cubes collected from 22 different patients. Their hyperspectral acquisition system acquired intraoperative data containing 826 successive spectral bands within the wavelengths of 400\,nm to 1000\,nm, with a spectral resolution of 2-3\,nm.
Preprocessing of the hypercube data was performed as outlined in the paper~\citep{fabelo2019helicoid}.

\cite{hyttinen2020oral} provide the second hyperspectral dataset used in this work.
The Oral and Dental Spectral Image Database (ODSI-DB),
is a larger dataset containing 316 different oral and dental hyperspectral images.
The hyperspectral images acquired in this dataset are from two different cameras. 171 out of the 316 images were acquired using a Specim IQ (Specim, Spectral Imaging Ltd., Oulu, Finland) line-scan camera, which has a spatial resolution of $512 \times 512$ and a spectral range of 400-1000\,nm with 204 spectral bands captured in total. The remaining images were obtained with the Nuance EX (CRI, PerkinElmer, Inc., Waltham, MA, USA) spectral scan camera, with a higher spatial resolution of $1392 \times 1040$ but fewer spectral bands. It features 51 bands ranging from 450 to 950\,nm. Due to their higher spectral resolution, in this work, only the line-scan (Specim IQ) hyperspectral images were selected for synthetic snapshot image generation. Denser spectral information is indeed beneficial for sampling of sensor responses during our image generation process. The hyperspectral data in this dataset comes preprocessed with flat-field correction from a blank reference sample, therefore white-balancing is not necessary for the ODSI-DB.

\subsection{Image Generation and Training Pipeline}
\begin{figure}
\centering
\includegraphics[width=\textwidth]{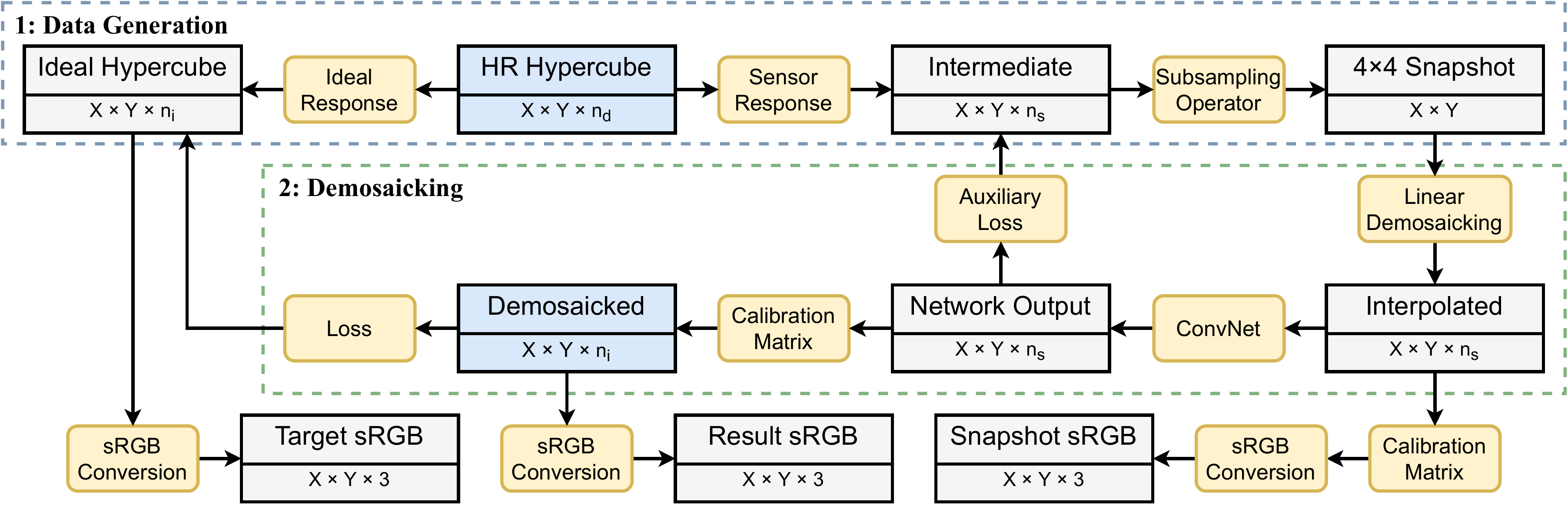}
\caption{Diagram of the hyperspectral snapshot image demosaicking algorithm simulated using high-resolution line-scan data. The rectangular boxes contain the types of data as well as their corresponding shape, whereas the rounded boxes show the operations in each step. The blue boxes indicate the input and output of the algorithm.} \label{demosaicking-framework-figure}
\end{figure}
\cref{demosaicking-framework-figure} illustrates the pipeline of the demosaicking algorithm of hyperspectral snapshot images. The entire framework consists of two parts.
The first part detailed in \cref{sec:generation} focuses on the generation of synthetic snapshot image and ideal high-resolution image datasets from high-resolution images (HELICoiD or ODSI-DB). 
The second part detailed in \cref{sec:training} involves the supervised learning method to get high quality hypercube reconstruction result. 

\subsubsection{Synthetic Image Generation Process}\label{sec:generation}
Synthetic image generation starts from a white-balanced high-spectral-resolution hyperspectral data cube (HELICoiD or ODSI-DB), referred to as HR Hypercube in the diagram.
We denote the size of a high-spectral-resolution hyperspectral cube as $X \times Y \times n_d$, where $X$ and $Y$ capture spatial and $n_d$ the spectral dimensions.

\myparagraph{Simulating the Spectral Response of the Snapshot Sensor}
Snapshot mosaic hyperspectral sensors only capture a discrete number of $n_s$ spectral bands with $n_s$ typically much smaller than $n_d$. For example $n_s=16$ for a $4 \times 4$ mosaic arrangement.
Each of the $n_s$ bands can have a non-trivial spectral response~\citep{pichette2017calibration} (e.g. bimodal and/or heavy tailed response) due to the parasitic effects such as harmonics, cross-talk and spectral leakage.
These responses are nonetheless typically calibrated in factory and can be retrieved from the calibration files of the camera sensor.

An intermediate high-spatial-resolution hyperspectral cube of size $X \times Y \times n_s$ can be generated by simulating the effect of camera sensor response on the high-spectral-resolution data.
More specifically, the intermediate hyperspectral cubes can be obtained by computing at each spatial location the inner products of the individual sensor responses with the high-resolution spectrum from the input data.

\myparagraph{Simulating the Spatial Response of the Snapshot Sensor}
Having simulated the spectral response and obtained a 
$X \times Y \times n_s$ intermediate hypercube, the final simulated 2D mosaic image can be derived by applying spatial subsampling as illustrated in \cref{hypercube-mosaic}(b).
More specifically, the hypercube is divided into smaller tiles with the same spatial size as the mosaic sensor array, and for each pixel in every individual tile, only one value from the $n_s$ wavelengths is preserved. Therefore, the synthetic snapshot mosaic image is a scalar-valued image of size $X \times Y$. 

\myparagraph{Simulating the Target Ideal Hyperspectral Data}
Given the non-trivial spectral response of
the captured $n_s$ spectral bands (harmonics, spectral leakage, cross-talk, etc.), a spectral correction matrix for snapshot systems provided by the camera manufacturer may reconstruct only a subset of $n_i \leq n_s$ spectral bands to ensure high-fidelity measurements for reconstructed bands. 

Resulting $n_i$ bands are designed to approximate \emph{ideal} sensor measurements by taking into account the response of ideal Fabry-P\'erot resonators~\citep{pichette2017calibration}. %
The corresponding optical band-pass response $f$ can be characterised as a Lorentzian function in optical frequency.
We express it in terms of the wavelength $\lambda$, centered around the central wavelength of each snapshot sensor $\lambda_0$, with full-width at half-maximum FWHM and with a quantum efficiency QE:
\begin{equation}
    f(\lambda;\lambda_0,\textrm{QE},\textrm{FWHM})=\textrm{QE} \frac{\alpha
    \lambda^2}{(\lambda-\lambda_0)^2+\alpha\lambda^2}
    \quad
    \textrm{with}
    \,
    \alpha = \frac{(\sqrt{\lambda_0^2 + \textrm{FWHM}^2} - \lambda_0)^2}
    {\textrm{FWHM}^2}
\end{equation} 

The number $n_i$ and characteristics $\lambda_0$, QE and FWHM of the ideal spectral bands are selected to capture all the reliable information contained in the $n_s$ spectral bands. These are typically provided by the camera manufacturer and are used to fit to the measured response curves.
A calibration matrix $C$ of size $n_i \times n_s$ to map the $n_s$ spectral measurements to the $n_i$ ideal spectral bands is also computed in factory and provided by the manufacturer~\citep{pichette2017calibration}.

While one could try to recover high-spectral-resolution data from low-spectral-resolution snapshot mosaic data, in many applications, it is sufficient to recover the spatial information lost by the spatial sampling process of the mosaic arrangement while estimating a reliable set of spectral bands.
As such, in this work, we aim to recover high-spatial-resolution information for each of the $n_i$ ideal spectral bands.
We refer to this target as the ideal hypercube in \cref{demosaicking-framework-figure}.
It can be estimated from the HR hypercube input data by applying ideal Lorentzian responses to it.
The size of the target ideal hypercube is thus $X \times Y \times n_i$.

\subsubsection{Learning for Demosaicking, Spectral correction and RGB Generation}\label{sec:training}

\myparagraph{Supervised Training Approach for Super-resolved Demosaicking}
The synthetic data generation in \cref{sec:generation} provides
paired high-spatial-resolution ideal hypercubes and 2D snapshot mosaic images.
Having access to such datasets,
we exploit supervised learning approaches to develop a demosaicking approach therefore achieving super-resolution of the captured mosaic data.

As outlined in the blue box in \cref{demosaicking-framework-figure}, the algorithm starts with a simple bilinear-interpolation-based demosaicking of the snapshot mosaic images.
This operation involves grouping the pixels inside the snapshot images according to the position of the sampled spectral bands, and then using bilinear interpolation along the X and Y axes to upsample each spectral band back to the original sensor size.
The resulting interpolated data is of size $X \times Y \times n_s$, i.e. the same size as the intermediate high-spatial-resolution hypercube.
While linear interpolation recovers the snapshot data to its original shape before subsampling, the resulting images can still look blurry.
It is now well established that deep learning can effectively refine image details with a fast inference speed~\citep{lugmayr2020ntire}, at least when applied to RGB data.
In our algorithm, a U-Net \citep{ronneberger2015unet} enhanced to accommodate residual units \citep{kerfoot2019lv} has been adopted for the super-resolution and demosaicking task.
The network contains a contracting path with 4 downsampling layers and 2 residual blocks at each resolution, as well as a symmetric expanding path with skip connections.

Rather than directly predicting the target ideal hypercube, we simplify the training procedure and take advantage of the known correction matrix $C$.
For this purpose, the network aims at inferring the intermediate hypercube of size $X \times Y \times n_s$. As such, the network output
has a same size and spectral characteristics as the bilinearly interpolated input hypercube but achieves sharper details.

\myparagraph{Embedding Spectral Correction}
From the initial output hypercube of the network, we compensate for the parasitic spectral effects of the sensor by applying the correction matrix $C$ to each spatial location.
The size and spectral characteristics of the resulting hypercube matches the target ideal hypercube.
To train the network and associated spectral correction, we use a loss that captures the error between the inferred corrected hypercube and the ideal hypercube.

In order to provide additional guidance with intermediate supervision, an auxiliary loss between the intermediate hypercube inferred by the residual U-Net and the intermediate synthetic high-spatial-resolution hypercube is also added. 
The idea behind this auxiliary loss is that instead of directly
learning to refine the spatial resolution and compensate for the parasitic spectral effect,
the network can be guided to focus solely on the image super-resolution task.

In terms of the choice of loss functions, we investigated~2 sets of configurations. For L1 loss configuration, both the training loss and the auxiliary loss are set to L1 loss. 
For Perceptual loss configuration, the L1 auxiliary loss is replaced with the feature reconstruction loss component of the perceptual loss \citep{johnson2016perceptual} as this has been shown to enable improved super-resolution performance.

Denoting the (non-spectrally-corrected) output hypercube from the residual U-Net as $\hat{y}^s$, the intermediate high-resolution hypercube as $y^s$, the pre-trained loss network for the perceptual loss as $\phi$, and the ideal hypercube as $y^i$, then the total loss $\ell$ can be expressed as follows
where the weight factor $\gamma$ is set to 0.001 empirically:
\begin{equation}
    \ell(\hat{y}^s;y^s,y^i) = \| y^i - C\hat{y}^s \| + \gamma \| \phi(y^s) - \phi(\hat{y}^s) \|^2
\end{equation}

\myparagraph{sRGB reconstruction}
For intuitive visualisation of the result, the linearly interpolated snapshot hypercube data, the demosaicked hypercube results and the ideal hypercube data that serve as the ground truth of the network are all converted into sRGB images. This is achieved by first converting the spectral data (corrected with $C$ where relevant) into CIE XYZ colour space using colour matching functions and assuming a D65 illuminant. We then convert the XYZ colour images into linear RGB colour space and apply gamma correction to obtain sRGB images.

\section{Results}
\begin{table}
\tbl{Quantitative analysis of the demosaicked hyperspectral cubes from the HELICoiD and ODSI-DB hyperspectral datasets. The second row (L1 and Perceptual) refers to the choice of the auxiliary loss in the algorithm. The 
ODSI-DB$\rightarrow$HELICoiD column indicates that the result is tested on the HELICoiD dataset using the network trained on the ODSI-DB dataset.}
{\setlength\tabcolsep{3pt}
 \begin{tabular}{lcccccc} \toprule
 %\multirow{2}{*}{\diagbox{Evaluation}{Model}}
 & \multicolumn{2}{c}{HELICoiD} & \multicolumn{2}{c}{ODSI-DB} 
 %& \multicolumn{2}{c}{Cross-Datasets} \\
 & \multicolumn{2}{c}{ODSI-DB$\rightarrow$HELICoiD} \\
 \cmidrule(lr){2-3} \cmidrule(lr){4-5} \cmidrule(lr){6-7}
 %\diagbox[font=\scriptsize]{Metric}{Loss}
 %Metric {\texttt{\textbackslash}} Loss
 %Eval$\downarrow$ {\texttt{\textbackslash}} Train$\rightarrow$
 %
 Eval {\texttt{\textbackslash}} Train
 %Loss
 & L1 & Perceptual & L1 & Perceptual & L1 & Perceptual \\ \midrule
 L1 & $0.081 \pm 0.036$ & $\mathbf{0.077 \pm 0.046}$ & $0.023 \pm 0.009$ & $\mathbf{0.018 \pm 0.007}$ & $0.248 \pm 0.104$ & $\mathbf{0.139 \pm 0.055}$  \\
 SSIM & $0.988 \pm 0.006$ & $\mathbf{0.989 \pm 0.005}$ & $\mathbf{0.998 \pm 0.001}$ & $\mathbf{0.998 \pm 0.001}$ & $0.946 \pm 0.022$ & $\mathbf{0.959 \pm 0.019}$ \\
 PSNR & $37.4 \pm 3.57$ & $\mathbf{38.3 \pm 4.25}$ & $47.1 \pm 3.26$ & $\mathbf{49.7 \pm 3.09}$ & $27.2 \pm 3.47$ & $\mathbf{33.6 \pm 3.03}$ \\ \bottomrule
\end{tabular}}
\label{metrics-results}
\end{table}
\begin{table}
\tbl{Perceptual metric scores on the sRGB images generated from the demosaicked hyperspectral cubes. The results were all tested on the HELICoiD dataset, and different demosaicking methods are compared (linear demosaicking, L1 model and perceptual model). }
{\begin{tabular}{lccccc} \toprule
 & \multicolumn{3}{c}{HELICoiD}
 %& \multicolumn{3}{c}{Cross-Datasets} \\
 & \multicolumn{2}{c}{ODSI-DB$\rightarrow$HELICoiD} \\
 \cmidrule(lr){2-4} \cmidrule(lr){5-6}
 %Metric
 sRGB Eval {\texttt{\textbackslash}} Train
 & Linear & L1 & Perceptual & L1 & Perceptual \\ \midrule
 L1 & $0.039 \pm 0.008$ & $0.012 \pm 0.004$ & $\mathbf{0.011 \pm 0.002}$ & $0.031 \pm 0.010$ & $\mathbf{0.023 \pm 0.007}$ \\
 SSIM & $0.932 \pm 0.021$ & $0.975 \pm 0.010$ & $\mathbf{0.979 \pm 0.008}$ & $0.928 \pm 0.026$ & $\mathbf{0.940 \pm 0.021}$ \\
 PSNR & $26.8 \pm 1.57$ & $34.8 \pm 1.83$ & $\mathbf{35.1 \pm 1.56}$ & $27.9 \pm 2.18$ & $\mathbf{29.7 \pm 2.16}$ \\
 LPIPS & $0.209 \pm 0.026$ & $0.130 \pm 0.018$ & $\mathbf{0.115 \pm 0.014}$ & $0.199 \pm 0.034$ & $\mathbf{0.178 \pm 0.036}$ \\ \bottomrule
\end{tabular}}
\label{srgb-results}
\end{table}
\textbf{Implementation Details.}
In our experiment, sensor information from Ximea xiSpec (MQ022HG-IM-SM4X4-VIS2) snapshot camera was used to simulate visible range (470-620\,nm) $4\times 4$ mosaic snapshot data. Synthetic image generation and demosaicking were performed on the HELICoiD and ODSI-DB datasets separately. The HELICoiD dataset contains 36 in vivo brain surface hyperspectral cubes in total, which we divided into 3 groups: 24 images for training, 6 images as the validation set and the other 6 images for testing. As for the ODSI-DB dataset, there are 122 hypercubes acquired from the line-scan sensor in total. 
78 hypercubes were used for training, 20 for validation and 24 for testing. As both datasets have cases where multiple hyperspectral data are obtained from the same subject, the dataset was split manually in order to avoid data from the same subject appearing in different groups. 

Both loss configurations described in \cref{sec:training} were tested in the experiment. For perceptual loss configuration, VGG-16 \citep{simonyan2015vgg} pre-trained network was used for feature extraction during the perceptual loss calculation, and the parameters of VGG-16 were fixed during training. In order to increase the number of training samples and limit the GPU consumption, the hyperspectral data were randomly cropped into smaller patches with a spatial size of $224 \times 224$. Random flipping and random multiples of $90^{\circ}$ rotation were also performed for data augmentation. The batch size was set to 3 for all training processes, and the evaluation losses are the same as the training losses. Adam optimization \citep{kingma2014adam} was used with the initial learning rate of 0.0001, and the best training models (lowest evaluation loss) after 10000 epochs with were selected for the proposed algorithm. 

\myparagraph{Quantitative Evaluation}
Three metrics have been used to evaluate the demosaicking results, including the average L1 error, the structural similarity index (SSIM) and peak signal-to-noise ratio (PSNR). The quantitative results of the demosaicked hyperspectral cubes from the HELICoiD and ODSI-DB datasets are listed in \cref{metrics-results}. In this table, results from both configurations with different auxiliary losses are shown, where the residual U-Net model with perceptual auxiliary loss performs slightly better than the model with L1 auxiliary loss, but the difference is subtle. However, when it comes to cross-dataset evaluation (ODSI-DB$\rightarrow$HELICoiD), where the model trained on the ODSI-DB dataset was used to directly test against HELICoiD's dataset without any fine-tuning, the perceptual loss model outperforms the L1 loss model significantly. It can also be observed that the cross-dataset results are slightly worse compared to the results of the model trained directly with the HELICoiD dataset, but it is still acceptable considering the domain gap between the HELICoiD and ODSI-DB datasets. 

\begin{figure}[tb!]
\centering
\includegraphics[width=0.9\textwidth]{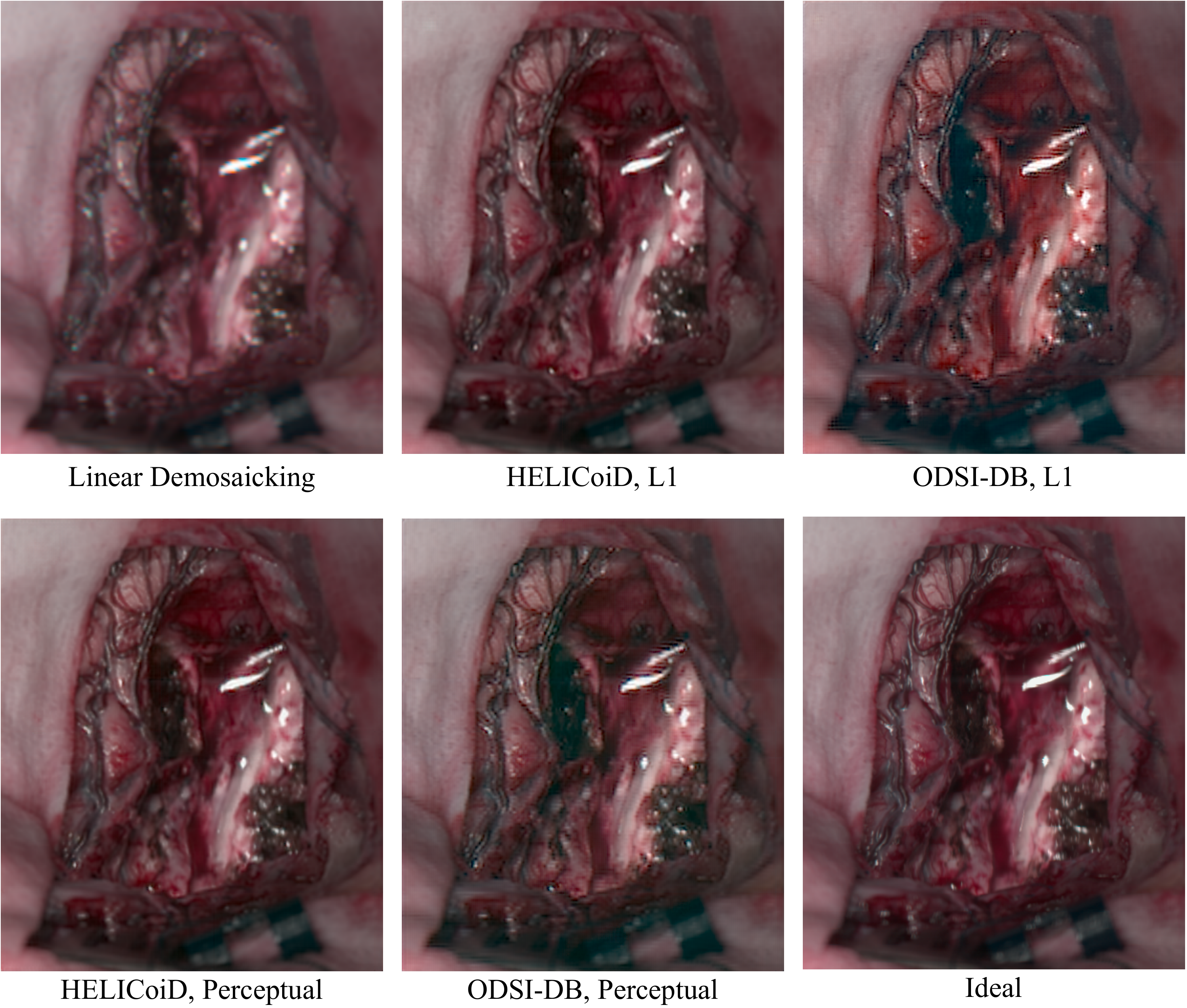}
\caption{Comparison between the sRGB images converted from the ideal hyperspectral cube and demosaicked results from linear interpolation and supervised learning models.} \label{demosaicking-result-srgb}
\end{figure}

The results were also evaluated based on the perceptual similarity of the sRGB images converted from the hyperspectral data. A perceptual similarity metric, namely LPIPS, was also used to simulate image comparison with human perception \citep{zhang2018lpips}. A lower perceptual score indicates the two images appear more similar to each other, with a score of 0 representing the best possible case, where the two images are the same. \cref{srgb-results} lists the perceptual scores of the sRGB images to evaluate the quality of the demosaicked hypercube data. Here the images for testing are all from the HELICoiD dataset, and the demosaicking model trained with ODSI-DB is not fine-tuned with any HELICoiD data. The demosaicking algorithm is also compared to the baseline linear demosaicking results, which is derived from the linearly demosaicked and spectral-corrected snapshot images that serve as the input of the residual U-Net as shown in \cref{demosaicking-framework-figure}. Similar trends can still be observed from this table, where the perceptual loss model outperforms the L1 model. Also, for the HELICoiD dataset in particular, the supervised learning based demosaicking algorithm achieves substantially better scores compared to linear demosaicking. One hyperspectral cube data from the HELICoiD test set has been selected to illustrate the result qualitatively, as shown in \cref{demosaicking-result-srgb}. The sRGB images show that the model trained on the HELICoiD datasets achieved respectable reconstruction results, with the result from the perceptual loss model having a slightly sharper image, which can be observed around the vessels as an example. On the other hand, the model trained on the ODSI-DB dataset can also recover the spatial resolution of the image to some extent compared to linear demosaicking, but it still suffers from artefacts as can be observed around the reflections in the image. 

\begin{figure}
\centering
\includegraphics[width=0.75\textwidth]{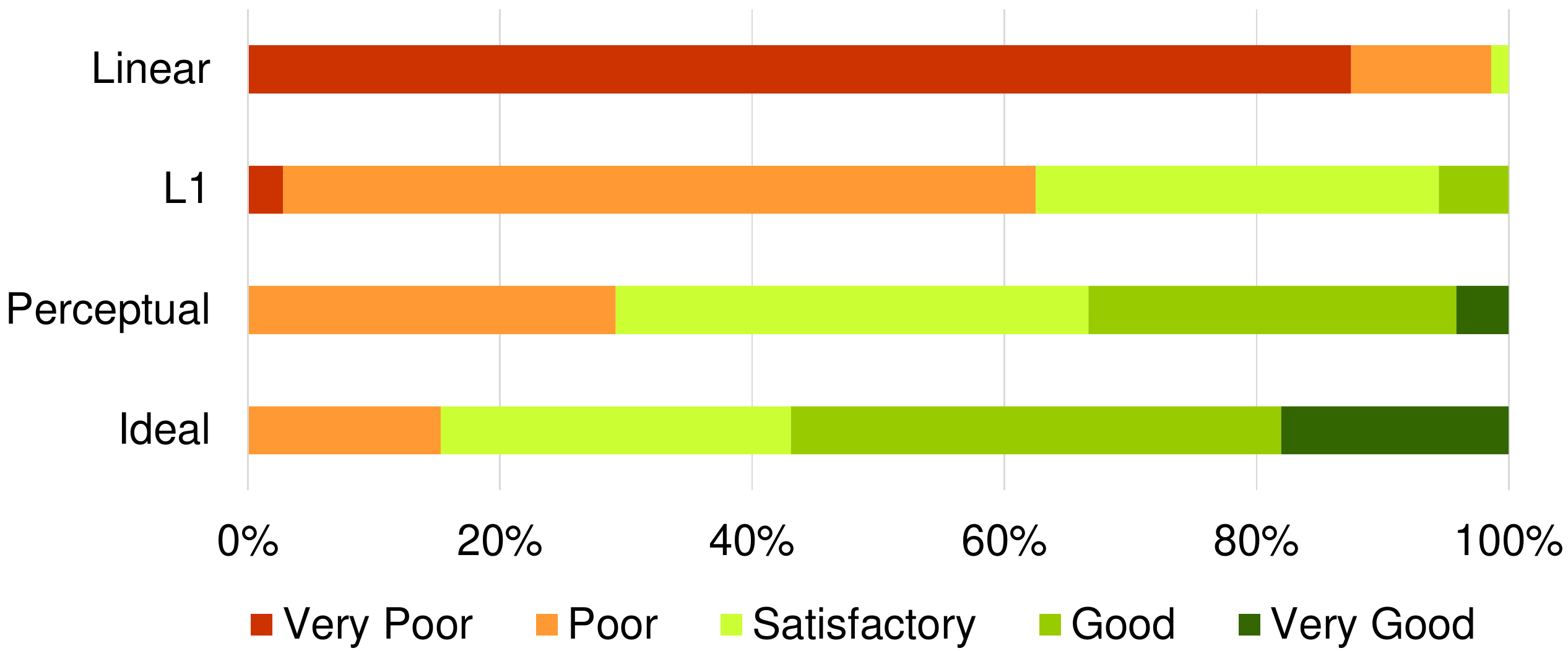}
\caption{Percentage distribution of image quality scores in Likert scale given by the clinical experts.} \label{survey-bar-chart}
\end{figure}
\myparagraph{User Study}Besides quantitative analysis of the data, a qualitative user study was conducted to evaluate the quality of the demosaicked images.
In this survey, the demosaicked HELICoiD test images were divided into 6 groups. Each group contained images with the same scene but generated from four different demosaicking methods, i.e. linear demosaicking, the proposed algorithm with L1 and perceptual losses, as well as the ideal demosaicked image. The images in each group were randomly shuffled and the label was hidden. 12 clinical experts were involved in the survey, who subjectively gave a Likert scale rating (integer score from 1 to 5, 5 is of best quality) to each image.
The quality scores from all experts are gathered and divided based on the demosaicking methods, and the percentage distributions are shown in the bar graph in \cref{survey-bar-chart}. The average score of all linearly demosaicked images is only $1.14 \pm 0.15$, and 2 experts claimed that some images seemed out of focus. The average scores for the proposed algorithm results from L1 and perceptual loss models are $2.40 \pm 0.41$ and $3.08 \pm 0.75$ respectively, indicating a higher image quality perceptually than linear demosaicking. The ideal demosaicked images achieve highest average score of $3.60 \pm 0.92$. We have also performed paired $t$-test between score statistics of linear demosaicking and L1 loss model, L1 loss model and perceptual loss model, as well as perceptual loss model and ideal demosaicking images, and the $p$-values are %$8.34\mathrm{e}{-24}$, $8.71\mathrm{e}{-9}$ and $2.71\mathrm{e}{-4}$ respectively, which are 
all smaller than the significance level of 0.05. This result indicates the differences of subjective image quality scores between different demosaicking methods are all statistically significant. 

\begin{figure}
\centering
\includegraphics[width=\textwidth]{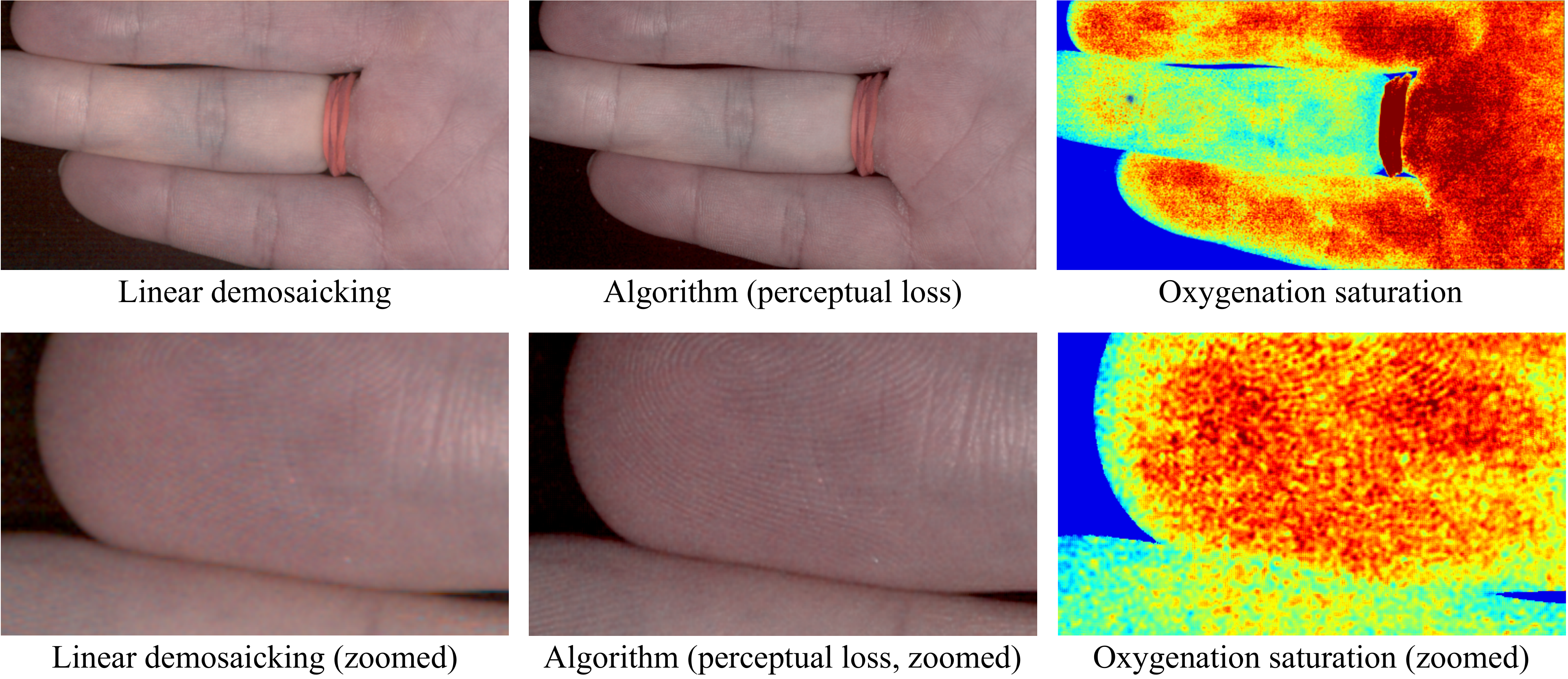}
\caption{Preliminary test results on real snapshot data converted into sRGB images. The linear demosaicking and the proposed algorithm are compared. The two images on the right also illustrate oxygenation saturation maps derived from hyperspectral information.} \label{demosaicking-result-real}
\end{figure}

\myparagraph{Preliminary Evaluation with Real Data}
One of the concerns regarding the supervised learning based demosaicking algorithm is that the entire framework relies heavily on synthetic data. Therefore, a real snapshot mosaic image of a hand captured by Ximea xiSpec (MQ022HG-IM-SM4X4-VIS2) was used to validate the effect of the algorithm, as illustrated in the converted sRGB images in \cref{demosaicking-result-real}. The difference between linear demosaicking and the proposed algorithm with perceptual loss model can be easily observed when we zoom in to closely investigate the details, where fingerprints can be recovered using the proposed algorithm. This result shows the generalisability of the algorithm, especially considering that the two models were never trained on real snapshot images. We also generated a blood perfusion map using the super-resolved hyperspectral data as shown in \cref{demosaicking-result-real} based on \citep{tetschke2016perfusion} to demonstrate the potential use of the algorithm in real medical applications. 
However, the cross-dataset results in \cref{metrics-results} and \cref{srgb-results} underline the domain gap in between different datasets, which may cause image artefacts.

\myparagraph{Real-time Performance}
We tested our prototype implementation on our computational workstation for clinical research studies (NVIDIA TITAN RTX 24GB, Intel Core i9 9900K) by taking advantage of Python, C++, OpenGL, Cuda, and Pytorch. 
The proposed algorithm achieved an overall processing time of approximately~45\,ms per $1088 \times 2048$ input image frame, including frame-grabbing, white balancing, bilinear demosaicking, followed by learning-based super-resolved demosacking, spectral correction and in the end either sRGB reconstruction or oxygenation saturation map estimation. The Pytorch-based U-Net super-resolution inference runs in~34\,ms.

\section{Conclusion}
In this work, we have proposed a hyperspectral snapshot image demosaicking algorithm for computer-assisted surgery using synthetic image generation and supervised learning. The simulated snapshot images and their corresponding ideal demosaicked images can be generated from publicly available hyperspectral image datasets acquired by line-scan sensors. A demosaicking framework has been developed with the adoption of a residual U-Net for hyperspectral image super-resolution, which can be trained with the synthetic image pairs. The quantitative and qualitative results show that the supervised learning approach is able to produce better reconstruction results compared to simple linear demosaicking, and it can still achieve a fast processing speed, which is beneficial for integration of the demosaicking algorithm into real-time surgical imaging applications. 
%Despite the convincing demosaicking result tested on a real snapshot image, the generalisability of the algorithm still remains to be investigated.
\revref{2}{1}
\revmod{
Future work includes further investigation on the generalisability of the algorithm when more real snapshot data are captured. As the proposed demosaicking approach separates the learning-based spatial super-resolution from spectral calibration, generalisation of our approach on real snapshot images can be expected, which has been demonstrated by the convincing results achieved with our preliminary real data evaluation. 
}
Besides, there is still room for improvements on speed and the image quality of the demosaicking algorithm. Nevertheless, the proposed demosaicking algorithm provides a solid step forward for medical hyperspectral imaging.

\paragraph*{Acknowledgments and Disclosures.}
This work was supported by core funding from the
Wellcome [203148/Z/16/Z], and EPSRC [NS/A000049/1].
This project has received funding from the European Union's Horizon 2020 research and innovation programme under grant agreement No 101016985 (FAROS project).
ME is supported by the Royal Academy of Engineering under the Enterprise Fellowship scheme [EF2021\textbackslash10\textbackslash110].
TV is supported by a Medtronic / RAEng Research Chair [RCSRF1819\textbackslash7\textbackslash34].
PL is funded by China Scholarship Council.
ME, SO, JS and TV are co-founders and shareholders of Hypervision Surgical.
TV holds shares from Mauna Kea Technologies.

\bibliographystyle{tfcse}
\bibliography{interactcsesample}

\end{document}